\documentclass[aps, preprint, PRL]{revtex4-1}
\usepackage{graphicx}
\usepackage{threeparttable}
\usepackage{subcaption}
\usepackage{subfig}
\usepackage{tabularx}
\usepackage{amsmath}
\usepackage{color}
\usepackage{amssymb}
\usepackage{gensymb}

\begin{document}


\title{A quantitative study of bias triangles presented in chemical potential space}
\author{Justin K. Perron}
\email{Perronjk@gmail.com}
\address{Joint Quantum Institute, University of Maryland. College Park MD}
\address{National Institute of Standards and Technology, 100 Bureau Dr. Gaithersburg, Maryland}
\author{M. D. Stewart, Jr.}
\address{National Institute of Standards and Technology, 100 Bureau Dr. Gaithersburg, Maryland}
\author{Neil M. Zimmerman}
\email{Neil.Zimmerman@nist.gov or by phone at +1 301 975 5887}
\address{National Institute of Standards and Technology, 100 Bureau Dr. Gaithersburg, Maryland}

\begin{abstract}
	We present measurements of bias triangles in several biasing configurations. Thorough analysis of the data allows us to present data from all four possible bias configurations on a single plot in chemical potential space. This presentation allows comparison between different biasing directions to be made in a clean and straightforward manner.  Our analysis and presentation will prove useful in demonstrations of Pauli-spin blockade where comparisons between different biasing directions are paramount. The long term stability of the CMOS compatible Si/SiO$_2$ only architecture leads to the success of this analysis. We also propose a simple variation to this analysis that will extend its use to systems lacking the long term stability of these devices.
\end{abstract}

\maketitle 

Generally, if someone wishes to determine the spin state of an electron in a solid state system they must employ some form of spin-to-charge conversion.  One popular method exploits the phenomenon of Pauli-spin blockade, where transport through series quantum dots is strongly influenced by the spin state of the electrons in the device.  There have been a number of impressive studies on Pauli-spin blockade in a variety of systems, including GaAs\cite{Johnson05}, SiGe\cite{Borselli11}, atomically precise phosphorous donor devices in Si\cite{Weber14}, and finally, Al\cite{Lai11} and Si\cite{Liu08} gated quantum dots.  In these studies the primary evidence for Pauli-spin blockade arises from the comparison of bias triangles in opposite current directions. However, the 3D nature of this type of data requires data from different bias directions to be plotted separately. This makes comparisons between data sets cumbersome.  Furthermore, triangles measured in different biasing configurations are shifted in voltage space. This is due not only to the capacitive coupling between the leads and the quantum dots but also any charge instabilities or drifts that occur between measurements. For these reasons, comparison of the shapes and sizes of bias triangles in previous studies have not been straightforward.  As a precursor to studies on Pauli-spin blockade we present measurements of bias triangles.  In addition to the usual positive and negative drain bias configurations we also present measurements in the positive and negative source configurations.  The absence of Pauli-spin blockade allows a straightforward interpretation of the data. Thus, we use these measurements to validate a new method of analysis.   Our analysis corrects for the coupling of the leads to the dots allowing for a much cleaner presentation of data from all four biasing setups on a single plot. This makes comparisons between biasing directions much simpler.

\begin{figure}
 \centering
		\includegraphics[width=0.95\linewidth]{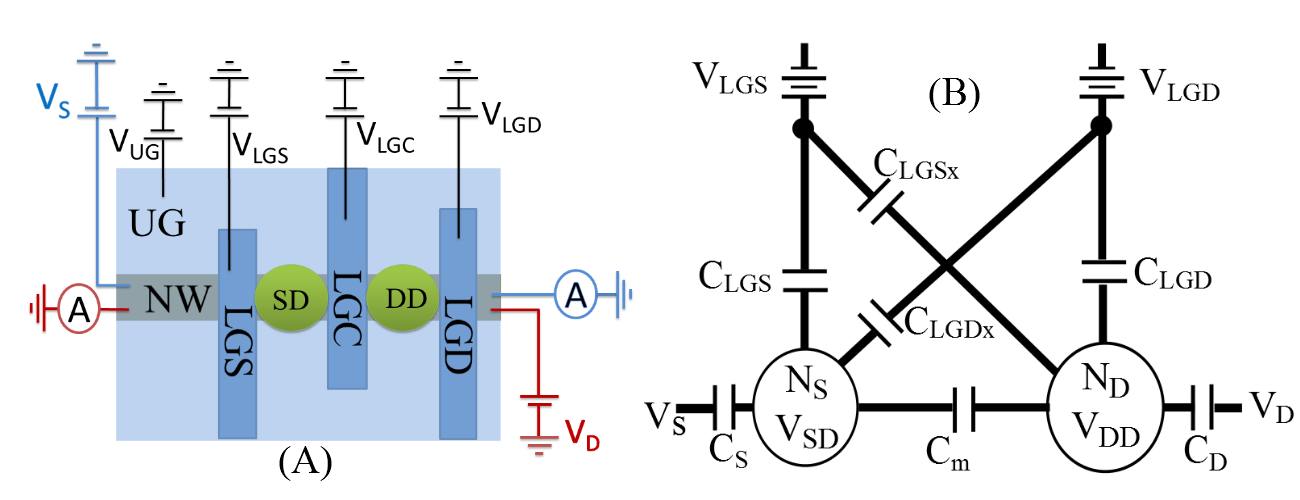}
		\caption{Experimental details and capacitive model a) Schematic of our device layout. Conduction occurs in a single crystal Si nanowire. This is allowed when a positive voltage is applied to the upper gate (UG) thus inverting the Si nanowire. Dots are formed between lower gates (LGS, LGC, LGD) by applying negative voltages to these gates creating local conduction barriers. The source dot (SD) and drain dot (DD) formed in the double dot measurements are shown in green.  Also shown is a schematic of our circuit. In one configuration (red) current is measured with a preamplifier connected to the source lead of the nanowire while a bias voltage, $V_D$, is applied to the drain lead.  In the other configuration (blue) a bias voltage, $V_S$, is applied to the source and current is measured at the drain. b) The capacitive model of our device. The lower gates nearest to the source (LGS) and drain (LGD) are capacitively coupled to the dots via $C_{LGS}$, $C_{LGD}$, $C_{LGSx}$ and $C_{LGDx}$. The upper gate (UG) and center lower gate (LGC) are also coupled to both dots (not shown). Finally the drain and source leads are coupled to the drain and source dots through $C_D$ and $C_S$ respectively.}\label{device}
\end{figure}

All measurements discussed in this manuscript were made with the device mounted on the mixing chamber stage of a dilution refrigerator with a base temperature of about $50~\rm{mK}$. The double quantum dot was formed in a CMOS-compatible Si nanowire. A schematic diagram of the device is shown in figure~\ref{device}a.  The single crystal nanowire is formed by mesa etching a 40~nm wide wire in the 30~nm thick Si layer of a silicon on insulator wafer. A high quality SiO$_2$ gate oxide is grown before depositing a gate layer of phosphorous doped polysilicon. This layer is etched into three 40~nm long conformal finger gates aligned perpendicular to the nanowire and spaced 40~nm edge-to-edge. These gates are used to electrostatically create conduction barriers in the nanowire as well as control the dot chemical potentials.  A second high quality SiO$_2$ layer is then formed on top of which a second gate layer of doped polysilicon is deposited. This layer forms a global upper gate that is used to invert the Si nanowire to allow conduction between n$^+$-doped ohmic contacts located microns away from the active device area.

Figure~\ref{device}a also shows a schematic of the circuits used to measure the device.  The black portion of the circuit is used in all measurements. The red circuit is connected when biasing the drain lead and is replaced with the blue circuit for measurements with source biasing. In order to measure the multiple biasing configurations necessary for this study, these two circuits were switched repeatedly. It is common to see uncontrolled changes in nano-device behaviour when changing the circuit configuration. The remarkable circuit stability across these changes is crucial to the success of the analysis discussed later. 

To model the device we use the capacitance diagram in figure~\ref{device}b. All of the capacitances used in this model are measurable with the device setup in various single quantum dot configurations. The results of these measurements are given in table~\ref{capacitances} (upper row); the capacitance values in this row result from measurements taken with various values of the gate voltages. Also listed are gate capacitances determined from a double dot charge stability measurement\cite{VDW02}. There is reasonable agreement between the gate capacitances, $C_{LGS}$ and $C_{LGD}$ and these values are also consistent with a more comprehensive study on these types of devices\cite{Thorbeck12}. However, there is a significant difference in the value of the coupling capacitance $C_m$. We speculate that this is due to the influence of the LGD and LGS gates on the barrier under LGC. To measure $C_m$, the device is configured to operate as a single dot between either LGD and LGC or LGS and LGC. During this measurement, the voltage on the lower gate not used in forming the dot is roughly 1.7~V above the value applied in double dot measurements. This voltage difference not only removes the barrier underneath said gate but also weakens the barriers formed under adjacent gates and therefore leads to a higher value of $C_m$.

\begin{table*}[t]
	\centering
	\begin{threeparttable}[b]
	\caption{Measured Capacitances in aF}
	\begin{tabular}{l|c c c c c c c}
		\hline \hline
		Measurement &~$C_S$~&~$C_m$~ &~$C_D$~ &~$C_{LGS}$ ~&~$C_{LGD}$~ &~$C_{LGSx}$ ~&~$C_{LGDx}$ ~\\
		\hline
		Single dot & 7.8\textsuperscript{\ddag} & 9.9\textsuperscript{\ddag} & 12.2\textsuperscript{\ddag} & 2.7\textsuperscript{\textdagger} & 2.3\textsuperscript{\textdagger} & 0.09\textsuperscript{\textdagger} & 0.05\textsuperscript{\textdagger} \\
		Double dot & NM & 5.2\textsuperscript{\ddag} & NM & 2.5\textsuperscript{\textdagger} & 2.3\textsuperscript{\textdagger} & NM & NM \\
	\end{tabular}
	\label{capacitances}
	\tiny
	\begin{tablenotes}
	\item[]{Representative uncertainties (one $\sigma$) are 10\% and 40\% for values denoted with \textsuperscript{\textdagger} and \textsuperscript{\ddag} respectively. These are determined from our estimated uncertainty in fitting to Coulomb diamonds, triangles and single gate curves.  Double dot measurements do not allow us to independently address all the capacitances of our model, therefore some values are listed as not measurable (NM).}
	\end{tablenotes}
\end{threeparttable}
\end{table*}

\begin{figure}
	\includegraphics[width = 0.95\textwidth]{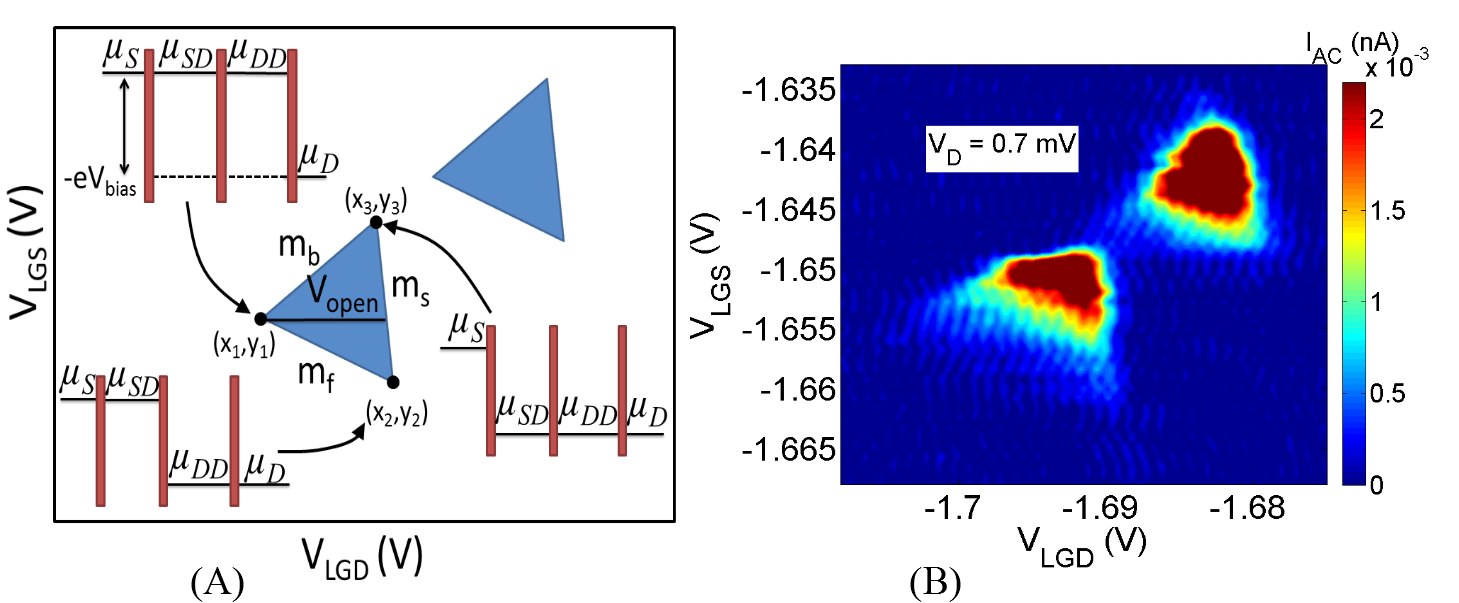}
	\caption{A) Schematic of bias triangles in the $V_D > 0$ configuration (red circuit in figure~\ref{device}a). The chemical potentials at the vertices of the electron triangle for a positive drain bias setup are also shown. The size of the triangles are determined by the magnitude of the bias via $V_{open}$ (see text). B) False color plot of bias triangles formed with 0.7~mV drain bias. The qualitative behavior described in the schematic of a) is observed in this data. Data was measured with $V_{UG}~=~2.5~{\rm{V}}$, $V_{LGC}~=~-1.1~{\rm{V}}$ and $V_{D,AC}~=~50~\mu{\rm{V_{pp}}}$ }\label{figure2}
\end{figure}

The chemical potentials of the dots are defined as $\mu_{SD}(N_S,N_D)=E(N_S,N_D) - E(N_S-1,N_D)$ for the source dot and $\mu_{DD}(N_S,N_D)=E(N_S,N_D) - E(N_S,N_D-1)$ for the drain dot.  Using the capacitive model of our system we determine the chemical potentials of the source and drain dots with $N_S$ and $N_D$ electrons respectively as

\begin{align}\label{musd}
	\mu_{SD}(N_S, N_D) &= \left(N_S -\frac{1}{2}\right)E_{SD} + N_D E_{C_m} \notag \\
	&-\frac{E_{SD}}{e}\left(C_{LGS}V_{LGS} + C_{S}V_{S} + C_{LGDx}V_{LGD}\right) \notag \\
	&-\frac{E_{C_m}}{e}\left(C_{LGD}V_{LGD} + C_{D}V_{D} + C_{LGSx}V_{LGS}\right)
\end{align}
\noindent and
\begin{align}\label{mudd}
	\mu_{DD}(N_S, N_D) &= \left(N_D -\frac{1}{2}\right)E_{DD} + N_S E_{C_m} \notag \\
	&-\frac{E_{C_m}}{e}\left(C_{LGS}V_{LGS} + C_{S}V_{S} + C_{LGDx}V_{LGD}\right) \notag \\
	&-\frac{E_{DD}}{e}\left(C_{LGD}V_{LGD} + C_{D}V_{D} + C_{LGSx}V_{LGS}\right).
\end{align}
\noindent Here $E_{SD}$ and $E_{DD}$ are the dot charging energies given by
\begin{equation}
E_{SD} = \frac{e^2 C_{DD}}{C_{SD}C_{DD}-C_m^2}
\end{equation}
\noindent and
\begin{equation}
E_{DD} = \frac{e^2 C_{SD}}{C_{SD}C_{DD}-C_m^2}.
\end{equation}
\noindent The coupling energy, $E_{C_m}$ is defined as
\begin{equation}
E_{C_m} = \frac{e^2 C_{m}}{C_{SD}C_{DD}-C_m^2}
\end{equation}
\noindent with $e$ the electron charge, $C_{SD}$ and $C_{DD}$ the total capacitances of each dot\footnote{To allow simple comparison, the variables in the popular work of \textcite{VDW02} are mapped to this result in the following way, $C_{g1(g2)} \rightarrow C_{LGS(LGD)}$, $C_{L(R)} \rightarrow C_{S(D)}$, $N_{1(2)}\rightarrow N_{S(D)}$ and $\mu_{1(2)}(N_1,N_2) \rightarrow \mu_{SD(DD)}(N_S,N_D)$.}. When biasing the drain lead (red circuit in figure~\ref{device}a) $V_S$ is equal to the potential of the input to the current preamplifier (close to ground), as is $V_D$ when biasing the source lead (blue circuit in figure~\ref{device}a) These chemical potentials can be used to determine the equilibrium charge configuration of the double dot. This is the familiar ``gravitational picture'' where electrons fall to the lowest available chemical potentials. As described in \textcite{VDW02} the number of electrons on the dots, $N_S$ and $N_D$, are fixed for large regions of gate voltage space.  These occupation numbers only change when a gate voltage(s) is changed in such a way that the chemical potential of a dot is lowered below that of the leads $\mu_{S(D)}$ or that of the neighboring dot. These boundaries between stable charge regions can be experimentally mapped out by measuring a response sensitive to charge reconfigurations in the device, such as a nearby quantum point contact\cite{Hasko93} or single electron transistor\cite{Zimmerli92, Schoelkopf98} or, as recently shown, the dispersive response of one of the device gates\cite{Reilly13}. In this manuscript, however, we measured the transport through the device. This method is only sensitive to current through the double quantum dot which can occur via electron or hole transport. In electron transport the charge state of the device cycles through the $(N_S, N_D)\rightarrow (N_S+1,N_D) \rightarrow (N_S,N_D+1)\rightarrow(N_S,N_D)$ configurations while hole transport occurs through the transitions $(N_S+1,N_D+1)\rightarrow(N_S+1,N_D) \rightarrow (N_S,N_D+1) \rightarrow (N_S+1,N_D+1)$. Focusing on electron transport for simplicity, at zero bias these transitions are allowed only when the chemical potentials of both dots are resonant with $\mu_S$ and $\mu_D$. Specifically, $\mu_S = \mu_{SD}(N_S+1,N_D) = \mu_{DD}(N_S,N_D+1)  = \mu_D$.  These conditions are met at what are called triple points. When $V_D > V_S$ the resonant condition is relaxed and becomes $\mu_S \geq \mu_{SD}(N_S+1,N_D) \geq \mu_{DD}(N_S,N_D+1) \geq \mu_D$ (for opposite bias polarity the inequalities are reversed). These conditions describe triangular regions in gate voltage space where transport is allowed through the device. Figure~\ref{figure2} shows both a schematic of these triangular regions (panel A) and corresponding data taken with a drain bias of 0.7~mV (panel B).  The qualitative description given by the above conditions and shown in panel A is observed in the data of panel B. Although helpful for understanding the physics of transport, the conditions described above are not very useful for analyzing measured triangles and comparing them with theory. To perform our analysis we now focus on deriving the positions of the vertices of the bias triangles. 

\begin{figure}
 \centering
		\includegraphics[width=0.95\linewidth]{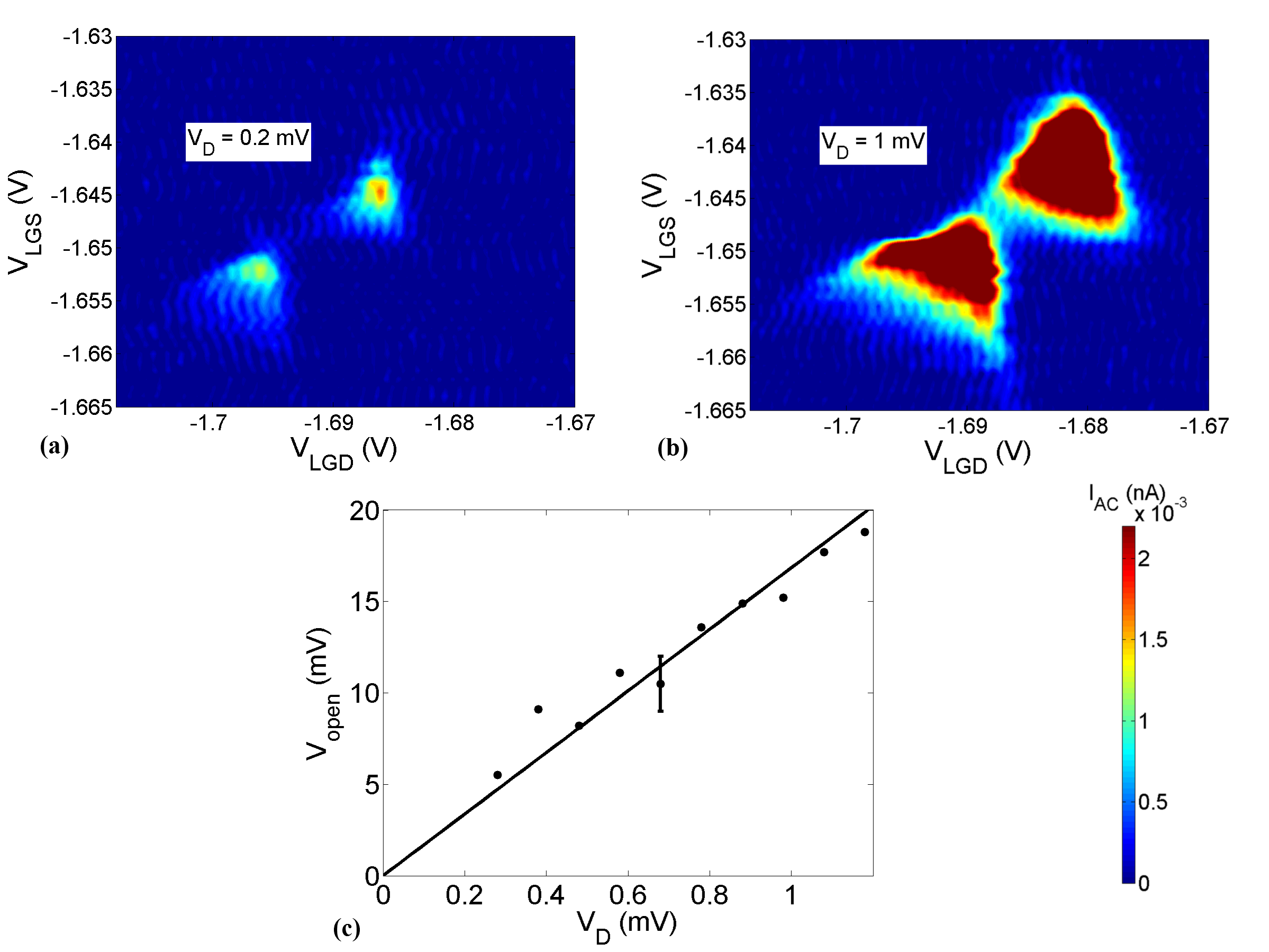}
		\caption{a) Bias triangles measured with $V_D = 0.2$~mV. b) Bias triangles measured with $V_D~=~1.0$~mV. There is an obvious increase in the size of the triangles with increasing bias voltage. c) A plot of $V_{open}$ determined from the data as a function of $V_D = V_D^{exp} - V_{offset}$ where $V_{offset}$ is the small offset voltage of the current preamplifier input. The vertical bar is a representative error bar. It is an estimate of how precisely we can select the edges of the bias triangles when determining $V_{open}$. The solid black line is determined using equation~\ref{openeq} and the capacitances from table~\ref{capacitances}. The agreement between the theory line, with no fitting parameters, and the data supports the use of our model. $V_{UG}~=~2.5~{\rm{V}}$, $V_{LGC}~=~-1.1~{\rm{V}}$ and $V_{D,AC}~=~50~\mu{\rm{V_{pp}}}$.}\label{Vopen}
\end{figure}

One vertex occurs when $\mu_{SD}(N_S+1,N_D)$ and $\mu_{DD}(N_S,N_D+1)$ are both resonant with the Fermi level of the lead connected to the current preamplifier (near ground potential).  This point is labeled as $(x_1,y_1)$ in figure~\ref{figure2}a and we refer to it as the $\mu = 0$ point.  From here the triangles open up with increasing $V_D$ along the $V_{LGD}$ axis by 
\begin{equation}
	V_{open} =V_D e^2/(E_{C_m}C_{LGDx} + E_{DD}C_{LGD}).
	\label{openeq}
\end{equation}
\noindent Figures~\ref{Vopen} a and b show data taken with $V_D = 0.2$ and 1.0~mV respectively.  The growth in the triangles with $V_D$ is clearly evident. Panel c shows $V_{open}$ determined from the data at multiple $V_D$ values.  Also shown is the dependence predicted from equation~\ref{openeq}. The agreement between data and the theory validates the use of our model.

Figure~\ref{figure2}a also shows the chemical potential conditions at the vertices of the bias triangles. By comparing these vertices we can determine the conditions that are satisfied along each side. The side we refer to as the base, with slope labeled $m_b$, has $\mu_{SD}(N_S+1,N_D)=\mu_{DD}(N_S,N_D+1)$ satisfied, while the other two edges, slopes labeled $m_f$ and $m_s$, have the conditions $\mu_{SD}(N_S+1,N_D)=\rm{constant}$ and $\mu_{DD}(N_S,N_D+1)= \rm{constant}$ respectively. Applying these conditions to equations~\ref{musd} and \ref{mudd} we determine the slopes of the triangle edges to be

\begin{align}
	m_b &= - \frac{(C_m - C_{SD})C_{LGD} + (C_{DD}-C_m)C_{LGDx}}{(C_{DD}-C_m)C_{LGS} + (C_m-C_{SD})C_{LGSx}} \label{mb} \\
	m_f &= -\frac{C_{LGD}C_m+C_{LGDx}C_{DD}}{C_{LGS}C_{DD} +C_{LGSx}C_m} \label{mf}\\
	m_s &= -\frac{C_{LGD}C_{SD}+C_{LGDx}C_m}{C_{LGS}C_m+C_{LGSx}C_{SD}}\label{ms}.
\end{align}

\noindent From these slopes, $V_{open}$ and a measurement of $(x_1,y_1)$ we can determine the coordinates of the other two vertices in this bias configuration. 

For the general case, all three vertices are determined by two effects: i) when the biasing condition is changed, the $\mu = 0$ point is shifted in gate voltage due to the capacitive coupling between the leads and the quantum dots. ii) Vertices $(x_2,y_2)$, $(x_3,y_3)$ move outwards from the $\mu~=~0$ point as $V_{open}$ grows.  The shifts from effect i) are derived from equations~\ref{musd} and \ref{mudd} to be
\begin{align}
	\Delta V_{LGD} &= - \frac{C_DC_{LGS} \Delta V_D - C_SC_{LGSx}\Delta V_S}{C_{LGD}C_{LGS}-C_{LGDx}C_{LGSx}} \label{shift1}\\
	\Delta V_{LGS} &= - \frac{C_SC_{LGD}\Delta V_S - C_DC_{LGDx}\Delta V_D}{C_{LGD}C_{LGS}-C_{LGDx}C_{LGSx}}\label{shift2}
\end{align}

\noindent with $\Delta$ denoting the difference between the measurement configurations. Using these shifts, and analysis similar to the $V_D>0$ case discussed above for effect ii), we can determine the coordinates of the bias triangle vertices in both drain-biased and source-biased configurations. These are given in table~\ref{vertices} where $V_{LGD}^*$ and $V_{LGS}^*$ are the experimental coordinates of the $\mu=0$ point in a single measurement (where $\Delta V_{LGD} \equiv \Delta V_{LGS} \equiv 0$), and where $V_{D(S)} = V_{D(S)}^{exp} - V_{offset}$ with $V_{offset}$ being the small voltage offset of the current preamplifier input.

\begin{table*}[h]
	\centering
	\begin{threeparttable}[b]
	\caption{Vertex coordinates in voltage space}
	\begin{tabular}{|c|c|c|}
		\hline \hline
		&Drain biased configuration& Source biased configuration \\
		\hline
		$x_1$& $V_{LGD}^* + \Delta V_{LGD}$ & $V_{LGD}^* + \Delta V_{LGD}$ \\
		$y_1$& $V_{LGS}^* + \Delta V_{LGS}$ & $V_{LGS}^* + \Delta V_{LGS}$ \\
		\hline
		$x_2$& $V_{LGD}^* + \Delta V_{LGD} +  V_D\frac{m_s\left(C_{SD}C_{DD}-C_m^2\right)}{\left(m_s-m_f\right)\left(C_{SD}C_{LGD}+C_mC_{LGDx}\right)}$ & $V_{LGD}^* + \Delta V_{LGD} +V_S \frac{\left(C_{SD}C_{DD}-C_m^2\right)}{(m_b-m_f)\left(C_mC_{LGSx}+C_{DD}C_{LGS}\right)}$ \\
		$y_2$& $V_{LGS}^* + \Delta V_{LGS} +  V_D\frac{m_f m_s\left(C_{SD}C_{DD}-C_m^2\right)}{\left(m_s-m_f\right)\left(C_{SD}C_{LGD}+C_mC_{LGDx}\right)}$ & $V_{LGS}^* + \Delta V_{LGS} + V_S \frac{m_b \left(C_{SD}C_{DD}-C_m^2\right)}{(m_b-m_f) \left(C_mC_{LGSx}+C_{DD}C_{LGS}\right)}$ \\
		\hline
		$x_3$& $V_{LGD}^* + \Delta V_{LGD} +  V_D\frac{m_s \left(C_{SD}C_{DD}-C_m^2\right)}{\left(m_s-m_b\right)\left(C_{SD}C_{LGD}+C_mC_{LGDx}\right)}$& $V_{LGD}^* + \Delta V_{LGD} + V_S \frac{\left(C_{SD}C_{DD}-C_m^2\right)}{(m_s-m_f) \left(C_mC_{LGSx}+C_{DD}C_{LGS}\right)}$\\
		$y_3$& $V_{LGS}^* + \Delta V_{LGS} +  V_D\frac{m_b m_s \left(C_{SD}C_{DD}-C_m^2\right)}{\left(m_s-m_b\right)\left(C_{SD}C_{LGD}+C_mC_{LGDx}\right)}$& $V_{LGS}^* + \Delta V_{LGS} + V_S \frac{m_s \left(C_{SD}C_{DD}-C_m^2\right)}{(m_s-m_f)\left(C_mC_{LGSx}+C_{DD}C_{LGS}\right)}$ \\
		\hline \hline
	\end{tabular}
	\label{vertices}
	\tiny
	\begin{tablenotes}
	\item[]{$\Delta V_{LGD}$ and $\Delta V_{LGS}$ are determined in equations~\ref{shift1} and \ref{shift2}. $V_{D(S)} = V_{D(S)}^{exp} - V_{offset}$ as discussed in the text.}
	\end{tablenotes}
\end{threeparttable}
\end{table*}

\begin{figure}
	\centering
		\includegraphics[width=0.85\linewidth]{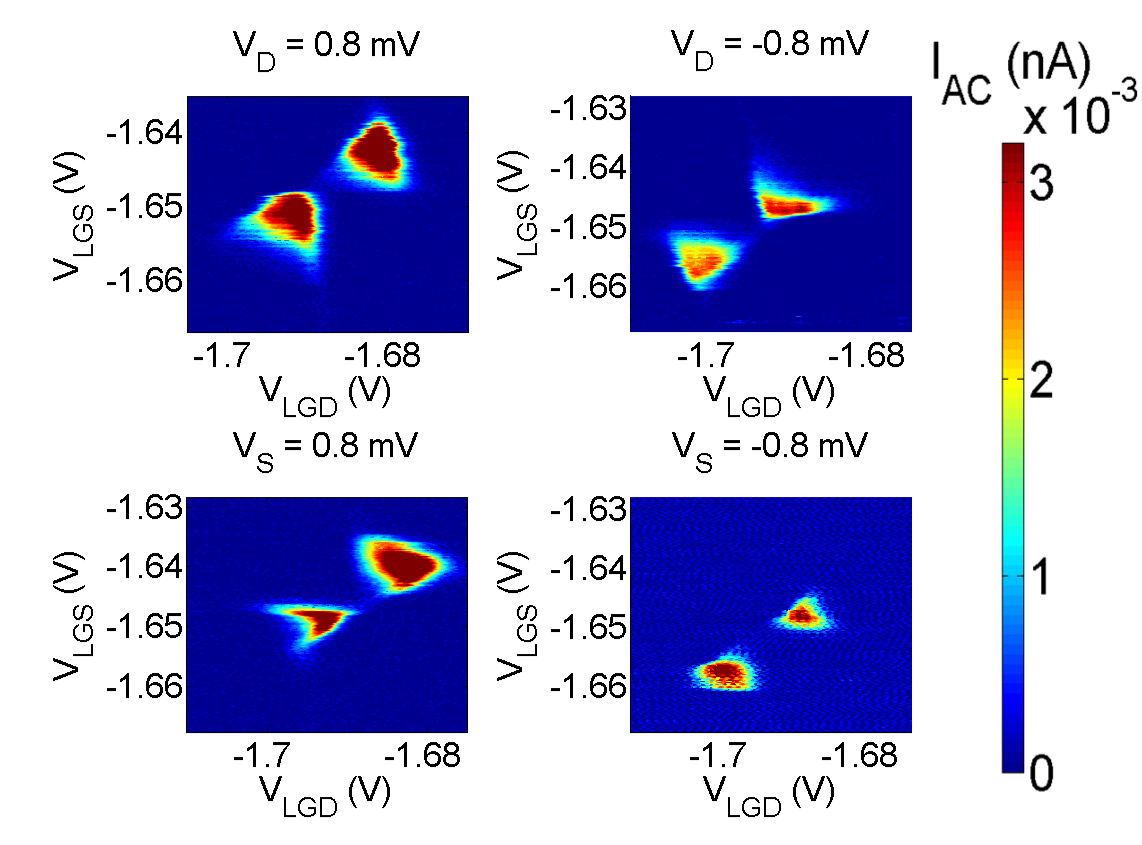}
		\caption{AC transport measurements in four different biasing setups, $V_D = \pm 0.8$~mV and $V_S = \pm 0.8$~mV. This presentation is typically how comparisons between different biasing setups are made.  In all cases $V_{LGC} = -1.1$~V, $V_{UG} = 2.5$~V and $V_{D,AC}~=~50~\mu$V.}
		\label{Falsecolours}
\end{figure}

Now that we have determined the predicted positions of the triangle vertices, we can move on to comparing bias triangles in different directions.  Figure~\ref{Falsecolours} shows AC transport data measured with $V_D=\pm~0.8~\rm{mV}$ and $V_S = \pm~0.8~\rm{mV}$.  Several of these triangles have some internal structure which is outside the scope of this paper; otherwise, they are qualitatively what is expected from theory. Triangles point in opposite direction depending on the direction of current flow, and in each bias setup there are two sets of triangles corresponding to the electron and hole transport\cite{VDW02}. Typically, false color plots like these are what is used for comparisons between bias configurations when analyzing Pauli-spin blockade.

The analysis resulting in table~\ref{vertices} provides a description of the triangle edges predicted by our model, which can be compared to the edges of measured triangles.  To perform this comparison we must determine the edges of the measured triangles in a systematic way. We achieve this by plotting a single contour at a current level just above the noise floor of our measurement. This corresponds roughly to the point where the appropriate chemical potentials are within $k_b T_e$ (where $T_e$ is the electron temperature and $k_b$ is the Boltzman constant) of the resonant condition which defines the edges in the theory. Therefore, the experimental contours constitute a slight overestimate of the triangle size. While we were unable to obtain an accurate measurement of $T_e$ due to structure in the zero bias Coulomb blockade peaks, we are confident that $T_e~<~1~K$. This corresponds to $\delta \mu \sim 0.1~meV$ or $\delta V_{gate} \sim 1.6 $~mV giving an upper bound on the overestimate. 

\begin{figure}
	\centering
	\begin{subfigure}{0.45\columnwidth}
		\includegraphics[width = 0.85 \columnwidth]{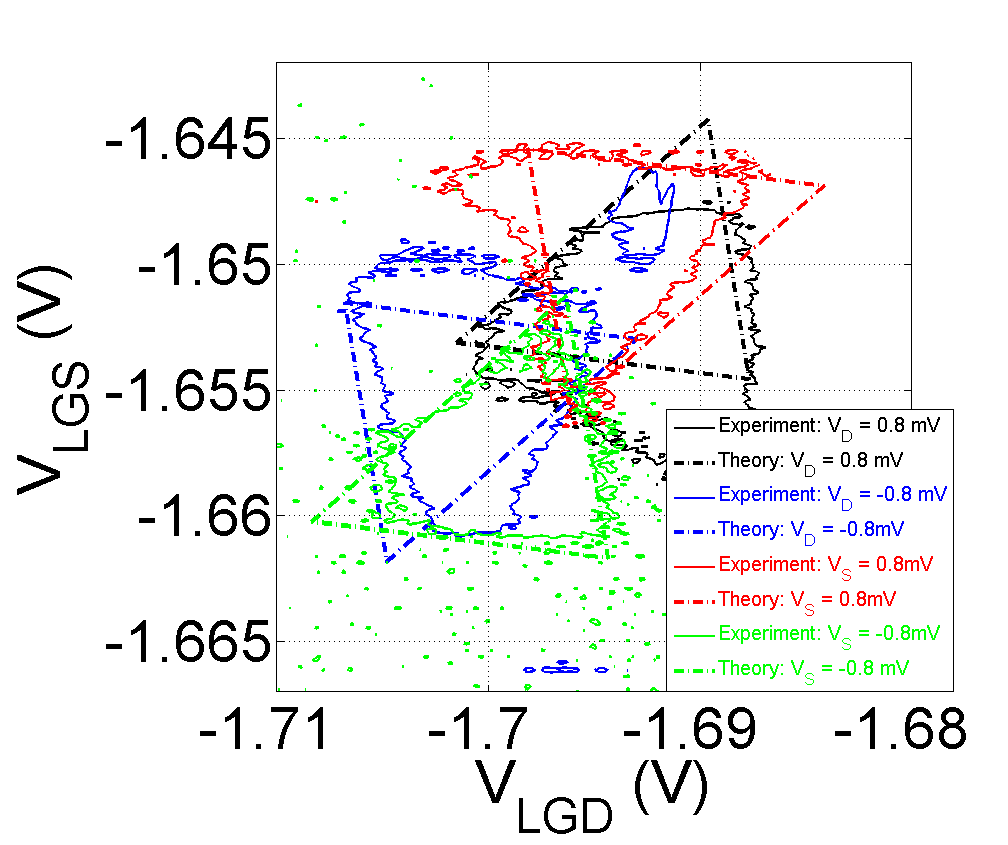}
		\caption{}
		\label{Vcontour}
	\end{subfigure}
	\begin{subfigure}{0.45\linewidth}
		\includegraphics[width = 0.85\columnwidth]{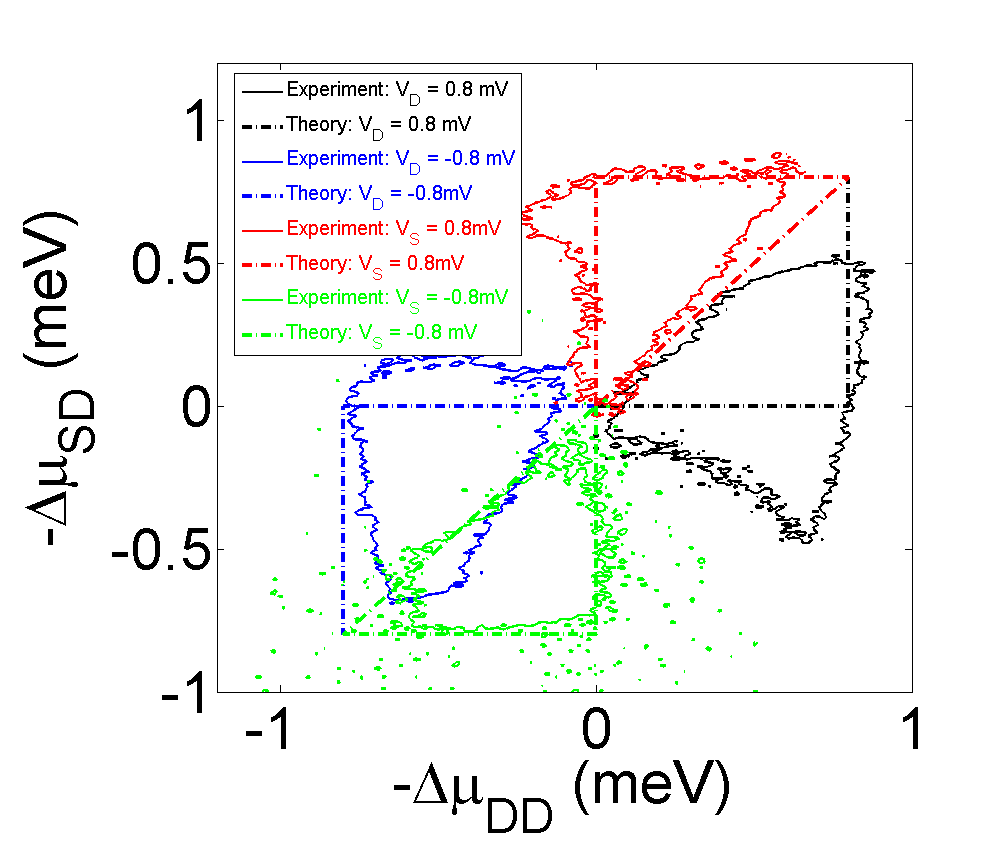}
		\caption{}
		\label{mucontour}
	\end{subfigure}
	\caption{a) Electron triangle data from figure~\ref{Falsecolours} plotted with a single contour at a level just above the noise floor in voltage space. Data from all four bias setups are shown. Also shown are triangles predicted using our capacitive model. Overlap between triangles from different biasing configurations makes data presented this way messy and comparisons between triangles difficult.  b) The same data converted to chemical potential space. The lack of overlap between triangles in chemical potential space leads to a much cleaner plot where comparisons can easily be made. Again the triangles predicted from theory are also plotted.}
\end{figure}

As mentioned above, false color plots similar to those in figure~\ref{Falsecolours} are what are typically used to make the comparisons that indicate Pauli-spin blockade. A much simpler comparison could be made if it was possible to display the relevant information from these figures on a single plot.  This is done in figure~\ref{Vcontour}, which shows the electron triangle data from all four biasing setups in figure~\ref{Falsecolours} using the single contour method. Also plotted are the triangle edges calculated using table~\ref{vertices}. When presented this way triangles overlap with each other making the plot much too busy for a simple comparison.  This overlap arises from two primary causes, the capacitive coupling of biasing leads to the quantum dots and charge offset drift $Q_0(t)$.

In devices using only high quality silicon-dioxide for dielectric layers, the issue of charge offset drift and charge reconfigurations is minimal\cite{Zimmerman01}.  This allows us, knowing the capacitances in table~\ref{capacitances}, to remove the overlap seen in figure~\ref{Vcontour}. We do this in figure~\ref{mucontour} by converting the plot from voltage space to chemical potential space where $\Delta \mu_{SD(DD)} = \mu_{SD(DD)}(V_{LGD},V_{LGS}) - \mu_{SD(DD)}(V_{LGD}^*,V_{LGS}^*)$. In this figure we have chosen to plot the negative of $\Delta \mu_{SD(DD)}$ so the triangles have similar orientation to those in figure~\ref{Vcontour}.  Theory predicts the bias triangles plotted this way should be right angle triangles extending in different directions from the $\mu = 0$ point. This greatly simplifies the predicted vertices, from the complicated equations of table~\ref{vertices}, to $x$ and $y$ values of 0 or $\pm eV_{D(S)}$. It should be noted that the theory lines in figure~\ref{Vcontour}, and the conversion of the data for figure~\ref{mucontour}, require only two input parameters, $V_{LGD}^*$ and $V_{LGS}^*$. Comparing the theory lines with the data we see reasonable agreement, in most places clearly better than the 40\% uncertainty associated with some of the input capacitances. To quantify this, we note that the $\mu=0$ vertices are all within 0.2~meV of the origin. This agreement equates to a charge offset drift, $Q_0(t)$, in our experiment of 0.03-0.04e, consistent with a separate measurement\cite{Zimmerman08} of $Q_0(t)$ interleaved with the bias triangle measurements (data not shown). This is a clear demonstration of the stability of not only the CMOS Si dots used in this measurement but also the measurement system. It is worth noting that circuit stability is far from guaranteed considering that, as discussed earlier, to apply bias voltages to opposite leads the circuit must be rewired between measurements.

Although the long term stability of our setup and devices allowed this analysis to succeed while neglecting $Q_0(t)$ completely, it is not a prerequisite for showing multiple bias triangles in a single chemical potential space plot. In systems where $Q_0(t)$ is more significant, its effects can be accounted for with separate $Q_0(t)$ measurements of the individual dots interleaved between bias triangle measurements. By interpolating the $Q_0(t)$ data to determine the appropriate values during the bias triangle measurements, one could correct for any drifts between measurements.

Apart from the aesthetic appeal of showing data from multiple biasing directions on a single graph, data plotted in chemical potential space would make the comparisons made when analyzing Pauli-spin blockade more straightforward. Data presented this way allow direct visual comparison of bias triangles in opposite directions; this is what constitutes the proof of Pauli-spin blockade. Furthermore, the conversion from voltage to energy necessary to quantify the energy splittings is performed prior to visualization of the data. This makes it substantially easier for people not intimately familiar with the analysis to interpret results. Also, discussions of series quantum dots generally rely upon the chemical potentials of the dots, not the applied voltages, and thus we feel data presented this way is more conducive to understanding.

Chemical potential space analysis similar to what is presented here may be useful in other systems as well, for example, in triple quantum dots, where a measured 2D voltage space can be very complicated\cite{Granger10}. In that case, it may be possible to work in a chemical potential space where a 2D plot would be neatly simplified by removing any contribution from motion of the chemical potential of the third dot. Secondly, compensation gating, where small voltages are applied to a second gate while sweeping or pulsing the gate of interest to prevent unwanted excursions in the chemical potential of a dot\cite{Veldhorst14, Yang11}, is easily understood using the chemical potential space picture. The type of analysis used in this paper provides a quantitative prediction of the magnitude of compensation voltages which may assist in experimental implementations.

In summary, we have measured bias triangles in a CMOS Si/SiO$_2$ series double quantum dot in four different biasing setups. By determining the location in voltage space of a single corner of one triangle, we were able to predict the position and shapes of subsequent triangles using a simple capacitive model.  With this model we converted the data from voltage space to chemical potential space where all four bias measurements of electron triangles could be clearly displayed on a single plot. This presentation will aid with the comparisons made in studies of Pauli-spin blockade. The analysis is performed using capacitances measured separately and only two input parameters.  Finally, we proposed instances where using this analysis to plot data in chemical potential space should be beneficial to the community.

\begin{acknowledgements}
	The authors would like to thank Garnett Bryant and Josh Pomeroy for their helpful discussions as well as Akira Fujiwara for providing devices. This work has been supported by an LPS grant ``Noise and coherent properties of silicon nanodevices''. 
\end{acknowledgements}

\bibliography{master}
\end{document}